# Foundations of Quantum Mechanics according to my teachers.

P. B. Lerner[1]

> "There are three reasons foundation of quantum mechanics are similar to sex:
>
> 1. Everyone considers herself/himself an educated amateur;
> 2. Professionals are treated with suspicion"
>
> And I forgot the third.[2]

**Introduction**

In my long and unhappy professional life, I had fortune to meet and listen to some of the titans of the 20th century quantum physics: Nobelists R. Peierls, H. Bethe, V. L. Ginzburg, S. Haroche as well as non-Nobel cult figures such as E. Teller, F. Dyson, Y. Aharonov or Ya. B. Zeldovich. But whatever insights and snippets were provided by their public lectures or overheard in private conversations, none of them concerned the foundations of quantum mechanics. Yet, meetings I had with the physicists of a lesser stature, sometimes in the context of my college studies and some—outside of it, had enlightened me with respect to this, deservedly suspicious (see epigraph) branch of speculative physics.

---

[1] Independent researcher; this treatise relies on personal recollection, so others could remember mentioned events differently. I am responsible for all the factual and conceptual errors. Excerpts from copyrighted work are reproduced under the "fair use" doctrine.

[2] Folk wisdom; to my recollection I heard it at the presentation in the seminar of Prof. I. Walmsley (now at Imperial College) at U. Maryland, Baltimore, 1993.

Act I. **Micromegas of Dr. D. A. Kirzhnitz** (1926-1998)[3]

Place: V. L. Ginzburg seminar on theory.

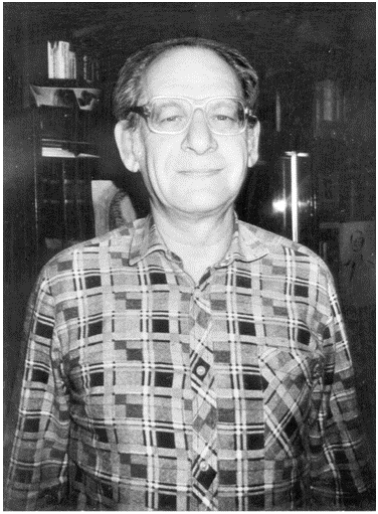

This photograph is not placed here as a formality. This is precisely how I remember him…

Nobelist V. L. Ginzburg for many years conducted a seminar on theoretical physics. There gathered the best physics minds from Moscow, environs and, frequently, from far away. A. D. Sakharov was a regular participant. A constant presenter on the current literature was D. A. Kirzhnitz, professor of his theory department at MFTI, my alma mater.[4]

K. "I always thought that a wave function was not a physical characteristic of an object but a representation of our knowledge of an object, something like a record of our knowledge about the object. For the imaginary being of cosmic size, the conventional Newtonian mechanics would be much like our quantum mechanics and for a microscopic sentient being, her quantum mechanics will be more similar to our classical one".

The allegory of the galactic animal is more transparent. Being limited to the scattering test of stars interacting with planetary systems, she would find a description in terms of scattering amplitudes more natural than our coordinates and momenta as a function of time. Moreover, because the trajectories of orbiting planets and their velocities would not be easily observable, and ruined after observation, they will be described in terms of stationary frequencies.

The opposite example of microscopic observer requires a little bit more imagination. At the first glance, the observer has to be of atomic size, which naturally obviates this example because then quantum mechanics has to be applied to its thinking process as well. But lo, the creature has to be only mesoscopic. Indeed, its size, for it to sample quantum trajectories directly, must be comparable to a typical atomic scattering length or coherence length in a superconductor.

---

[3] Micromegas is a short novel by Voltaire.
[4] Among the people who could be considered pupils of Prof. Kirznits, one can mention A. Linde (Stanford), one of the pioneers of Inflation, the theory.

Scattering lengths in the case of Feschbach resonances (or Ramsauer scattering) can be pretty large compared to atoms. For the mesoscopic observer, we can imagine the picture somewhat reminiscent to the Fig. 2-1 in Feynman and Hibbs [Feynman1965], or the representation of the quantum world similar to the representation of macroscopic reality provided to the bat by its echolocation.

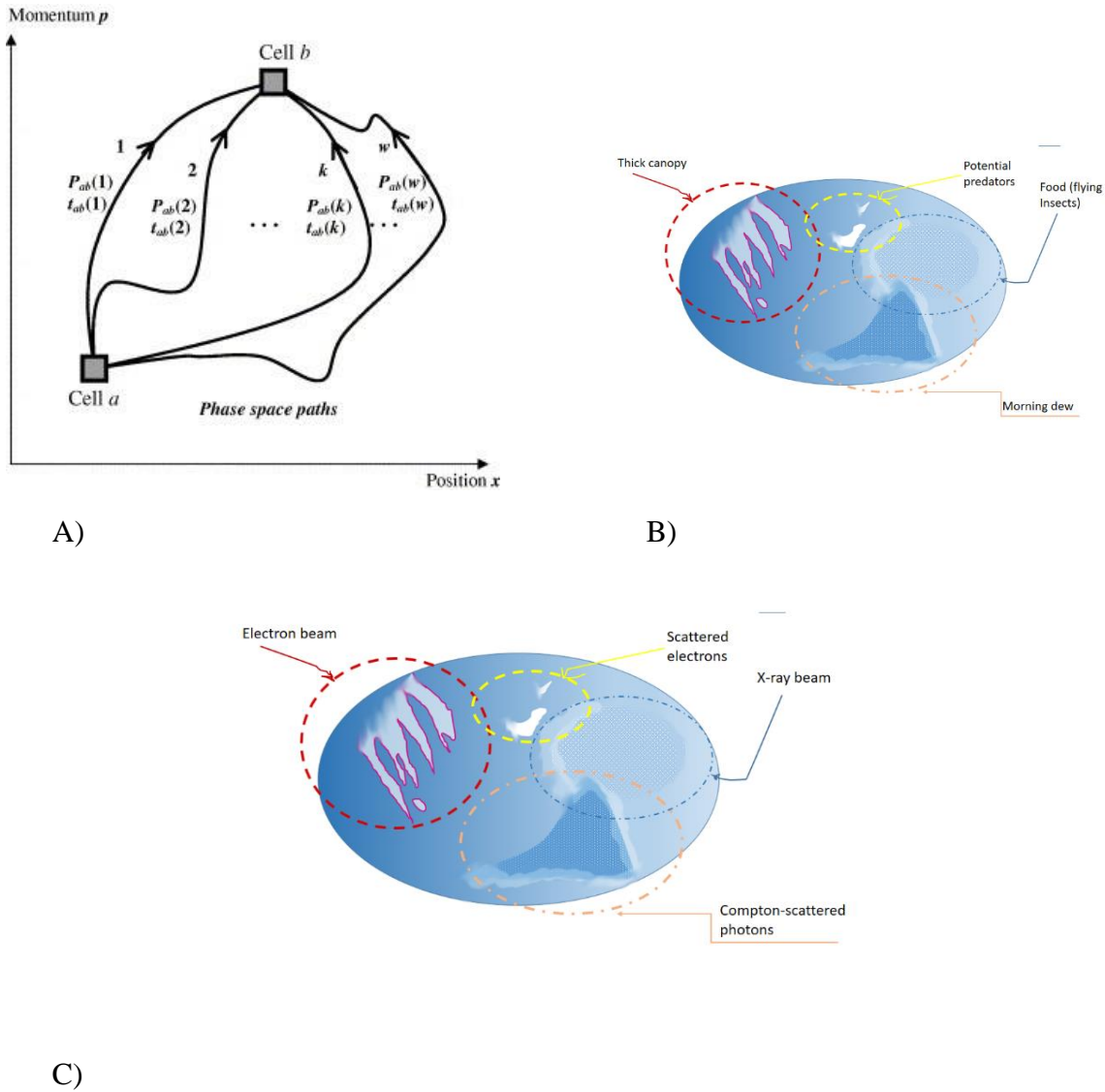

Fig. 1. A) Imagined quantum trajectories between vision cells of a mesocopic animal; B) The field of ultrasonic vision of the bat imagined by the author; C) Fictional mesocopic animal field of vision of electron-photon scattering. It observes X-ray photons through photoeffect in the "vision cells" and electrons through the Compton scattering of "soft" optical photons.

Act II. **K. A. Ter-Martirosyan** (1922-2005)

Place: Moscow subway

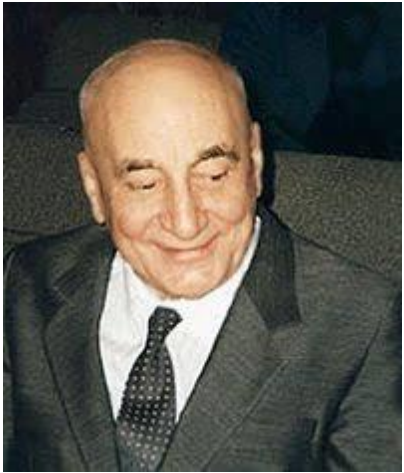

Ter-Martirosyan's younger years fell on the dawn of the Soviet nuclear program and he was not internationally famous. Yet, similarly to P. Ehrenfest two generations ago, he participated in the rearing of a number of world-caliber physicists: A. Voloshin, M. A. Shifman (U. Minn.) and my coursemate, late Ya. I. Kogan (U. Cambridge). Ter-Matrirosyan never taught me personally, but as a participant of his seminar, the pupil and friend of some of his students, I recollect this encounter.

The last place I met him was Moscow subway, in 1986 or 1987 where we returned home—we lived along the same line—after Rudolf Peierls delivered his lecture in Kapitsa Institute (Institute for Physical Problems) as a harbinger of beginning perestroika. I noticed to K. A. his accented but excellent Russian—he had Russian-born wife, Eugenia Kanegisser—a member of Landau's circle and the sister of executed poet and revolutionary terrorist Leonid Kanegisser.

On his question of my opinion of the lecture, I could not offer but a few laudatory platitudes. But Karen Aveticovich was nonplussed—Peierls talked about quantum properties of the particle—an old chestnut, "but the pertinent question now is the situation when the quantum properties of the measurement device start to manifest themselves".

This was the end of our conversation because we reached his station but the fact, that many modern approaches to the foundations of quantum mechanics: QBism, decoherence, Quantum Darwinism, consistent histories—do not include "reduction" or "collapse of the wavepacket" testifies to the prescience of his remark. We do not need to draw an artificial boundary between the classical and quantum worlds—on the opposite, we have to explain the emergence of the observable, approximately Newtonian world from quantum mechanics and the theory of relativity.

Intermission. **D. I. Blokhintsev** (1908-1979)
Place: My room reading his textbook

I did not study with D. I. Blokhintsev—he taught at Moscow State University—but I used his textbook extensively and socialized with his grad students.

The origin of Blokhintsev's views on foundations of quantum mechanics was ironic—after the Stalinist pogrom of Soviet biology in the aftermath of the Second World War, there were demands to apply Marxist-Leninist dogma to physics. Stalin was not supportive of these demands because he was informed that the imposition of "Marxist view" on physical sciences can threaten his nuclear bomb project. Yet, Dmitrii Blokhintsev as a true soldier of the Party responded to the call to reconcile quantum mechanics to "materialist" philosophy. What he produced was so ahead of his time that friend of Paul Dirac and Nobelist himself, Igor Tamm, who was a firm supporter of Copenhagen interpretation, exclaimed: "Why he writes all this nonsense? He could have known better".

Namely, Blokhintsev proposed that what is called "the reduction of the wave packet", namely the disappearance of the off-diagonal elements of the density matrix happens because of inevitable interaction of the observed quantum system with a measurement device. This point of view was later called "decoherence" and ascribed to H.-D. Zeh and V. Zurek.[5] Because both grew up close to the Soviet sphere of influence it is hard to imagine that both of them were completely unaware of Blokhintsev's textbook. [Blokhinstev1964, Blokhintsev1969]

I provide a photocopy of the pages from his book Quantum Mechanics, Select Problems [Blokhintsev88] to demonstrate, how close he came to decoherence. Namely, he produced a thought experiment with the measurement device consisting of a model potential (Fig. 2), in which a probe particle undergoes zero-point oscillations. When acted upon by an external particle, it slides from a very large potential barrier in the direction of momentum transfer. Because of large (i.e. "macroscopic") dimensions of the confining potential, the interference terms between right-moving and left-moving wavepackets are negligible. Henceforth, though the wavefunction of the probe particle is symmetric with respect to rightward and leftward motion—a natural feature of the wave equatons—because of large amplification of transferred momentum and a negligible tunneling matrix elements between the localized "left" and "right" states, the probe behaves practically as a Newtonian particle.

Further ahead of other versions of decoherence, Blokhintsev suggested from the beginning that interaction with a measurement apparatus could be enough in most cases, to provide decoherence, without attaching an external bath.[6]

---

[5] The reputation of Vojczech Zurek as a great Quantum Mechanic seems secure enough with his "no-cloning" theorem and establishing continuity between wave and particle properties in quantum mechanics—both papers co-authored with W. Wootters—and can only be enhanced by recognition of Blokhintsev's insights ahead of his time.

[6] This idea was inverted by the author in co-authorship with Paolo Tombesi (Phys. Rev. A, 1993, 47(5):4436-4440), in which measurement procedure itself was described as an external bath of thermally excited oscillators. To my great consternation, the editors did not correct my (awful, at the time) English and the paper did not receive acceptance it deserved.

Act III. **L. I. Ponomarev** (1937—)
Place: Ithaca, c. 2010

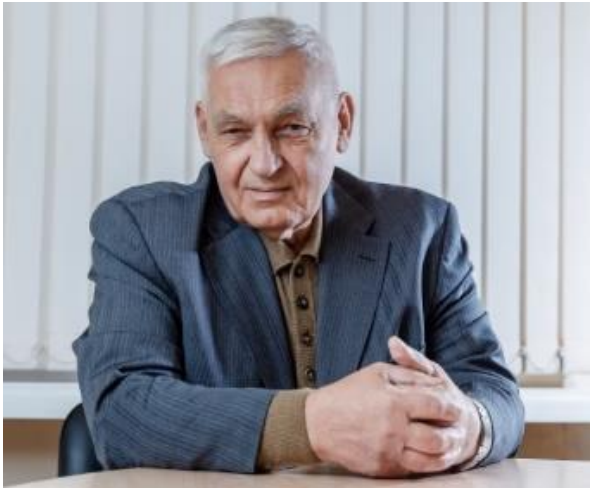

Leonid Ponomarev taught me quantum mechanics in the 1977-1978 school year. His outlook on the wavefunction (or density matrix) was pretty close to what we already discussed in the Abram Kirzhnits section—"the wavefunction as a notebook of our knowledge about quantum system".

This analogy was expounded on in his popularization "Under the sign of quantum" illustrated by the pen graphics by the author himself (see [Ponomarev2007] and Appendix B). Leonid Ivanovich was so kind as to send the copy of this wonderful book to me in my Upstate NY wilderness.

Namely, he compared a wavefunction—the analogy with density matrix seems more appropriate because of a two dimensional visual representation—as the record of a chess party that used to be published in general-audience magazines and newspapers in the last century.

He asserted that, for instance, we do not know many details of the sensational 1927 match between Capablanca and Alexander Alekhine. We will never know positions and coordinates of the air particles, nor the exact number of attendants, the location of chairs and so on. But we have a transcript of their chess moves and that is all information we need to reconstruct the same game, at any time, any place (see the copy in Russian).

I would go even further: even in our "classical" world we rarely use Newtonian coordinates and momenta to plan our own movements! Imagine that one plans a trip to Italy. It would be totally impractical to work with coordinates and momenta. Instead, we mark a notebook, or now a smartphone with the places of interest, the most convenient routes and train schedules, etc. etc.

Even an absence of instantaneous one-to-one correspondence of particle with its coordinates and moment has a classical analogy. Imagine one's "entanglement" with a wife. She has a conference in Verona and for the day X you planned to visit Roman amphitheater together after your return from trip to Bologna. But on the day X amphitheater was closed and your wife took a train to Bologna and bumped into you in one busy street before she informed you of the changing plans.

This analogy is not to suggest that our world, on some level, is Newtonian or Bohmian. But to find out that you and your wife are not quantum particles—what means that your de Broglie wavelengths are minuscule and you do not interfere (Ha! Ha!), or that your action is an astronomical multiple of $\hbar$–one needs some weird version of Bell's inequalities and, anyway, one cannot make statistically reproducible experiments with one's wife as scientists do with electrons. [Bell1987][7]

**Conclusion**

There is no conclusion for the debate about foundations of quantum mechanics in sight. In the view of this author, the Many Worlds interpretation of quantum mechanics, which replaced Copenhagen interpretation as the one accepted by doyens of physical sciences, first and foremost, the cosmologists—cannot currently provide an example of *experimentum crucis*—to allow for its certain refutation or uncertain support. A commendable effort in this direction by Rauchiger and Brenner (FR2018) does not seem to be inconsistent with standard quantum mechanics (Lerner2019). Thank you, my teachers!

---

[7]For canonical form of Bell inequalities, see e.g. Bell, 1987, Chapter 7. Also, compare my "wife paradox" with beginning of his lecture in Chapter 16.

**Appendix A**. Excerpts from Blokhintsev's 1988, providing more detailed rendering of the argument from his 1964 Principles of Quantum Mechanics. English version of similar material can be obtained from Blokhintsev, 1969.

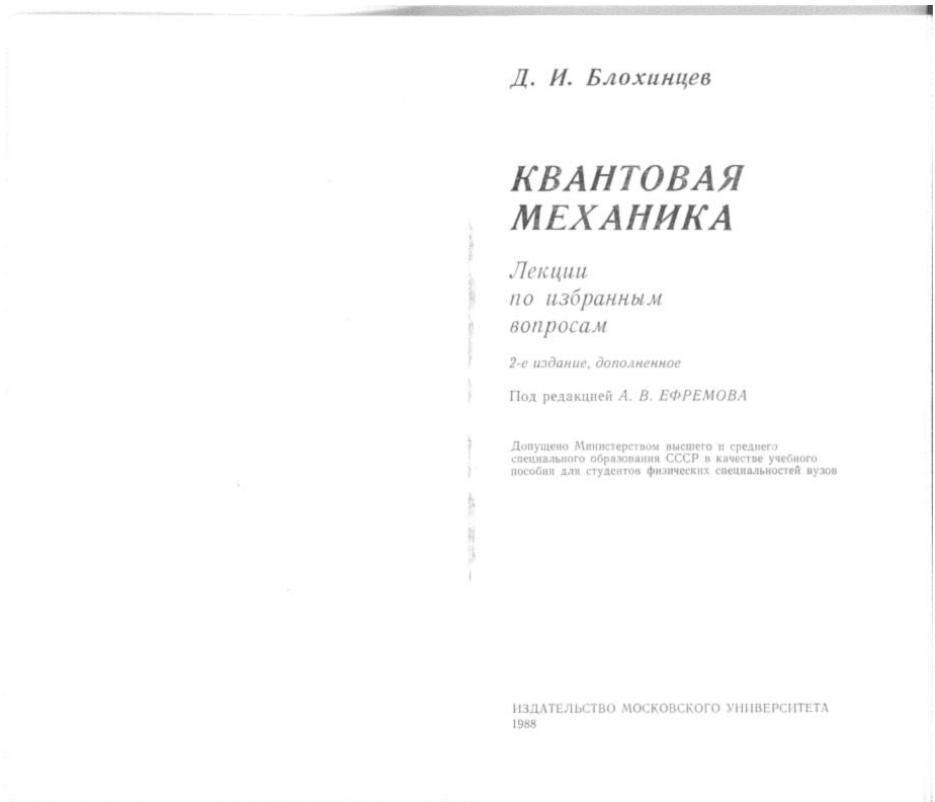

лагает ψ на право- (ψ₁) и лево- (ψ₂) поляризованные пучки. В этом случае пучки ψ₁ и ψ₂ остаются когерентными, и их можно вновь свести в когерентный пучок ψ. Детекторы $D_1$ и $D_2$, будучи макроскопическими системами, нарушают эту когерентность, более того, их роль обязательно сопровождается необратимыми процессами, увеличением энтропии. Это увеличение энтропии, как было разъяснено в лекции 9, есть плата за информацию.

В рассмотренных выше примерах исходная макроскопическая обстановка $M$ и вместе с ней сам исходный ансамбль $(M+\mu)$ задаются экспериментатором. Это, конечно, особый случай. Квантовые ансамбли существуют в природе и сами по себе, независимо от экспериментаторов. Они существовали и тогда, когда вообще никаких экспериментаторов не было на свете [1, 2]. Примером природного ансамбля является ансамбль космических лучей. Этот ансамбль определяется солнечной деятельностью и магнитным полем Земли. В данном случае экспериментатор ставит своей задачей выяснить природу ансамбля, уяснить состав частиц и спектр их энергий на входе лучей в атмосферу, т. е. определить $\hat{\rho}_M(0)$, и изучить дальнейшее развитие этого ансамбля, т. е. определить $\hat{\rho}_M(t)$. Время $t$ отсчитывается в этом случае высотой $H$, на которой изучается ансамбль. Частицы космических лучей имеют скорость, близкую к скорости света $c$, поэтому $t=(H-H_0)/c$, где $H_0$ — какая-нибудь заатмосферная высота (рис. 6). При этом изучаются разные экземпляры частиц $\mu$. Но все они объединены принадлежностью к одному квантовому ансамблю, описываемому статистическим оператором $\hat{\rho}_M(0)$. На рис. 6 это обстоятельство отмечено указанием на то, что из $N$ падающих первичных частиц $N_1$ используются для измерений на высоте $H_0$

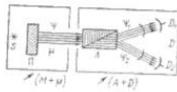

Рис. 5. Измерение в когерентном ансамбле. $S$ — источник частиц; $П$ — поляризатор (система $M+\mu$). Анализатор $A$ раскладывает исходный пучок на пучки ψ₁ и ψ₂ с различной круговой поляризацией; $D_1$, $D_2$ — детекторы частиц различной поляризации

(для получения информации о $\hat{\rho}_M(0)$), а $N_2$ частиц идут на измерения на высоте $H$ (для получения информации о $\hat{\rho}_M(t)$). Другим интересным примером природного квантового ансамбля является природный атомный реактор, обнаруженный в Габоне (Западная

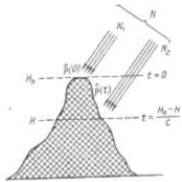

Рис. 6. Космические лучи исследуют на большой высоте $H_0$ в целях выяснения их первичного состава (определение $\hat{\rho}(0)$); изучая их на меньших высотах $H$, получают сведения об их развитии (определение $\hat{\rho}(t)$)

Африка), который существовал примерно два миллиарда лет тому назад и работал более полмиллиона лет [22].

Всюду, где в природе протекают квантовые процессы, мы имеем дело с квантовыми ансамблями. Такие ансамбли обычно принадлежат к числу некогерентных и открытых ансамблей.

### ЛЕКЦИЯ 12.
### ПРОСТЕЙШИЙ ПРИМЕР ВЗАИМОДЕЙСТВИЯ МИКРОЧАСТИЦ С ИЗМЕРИТЕЛЬНЫМ ПРИБОРОМ

Рассмотрим измерение, относящееся к микрочастице $\mu$, которая имеет массу $\mu$. Ее координату обозначим $x$ (для простоты ограничиваемся одним измерением); импульс — $k$. Предположим, что исходное состояние частицы является суперпозицией двух плоских волн:

$$\varphi_k(x)=A^+\exp(ikx)+A^-\exp(-ikx); \quad (12.1)$$

два частных состояния с $+k$ и $-k$ в этом случае когерентны.

Мы хотим узнать, каково направление движения частицы? Иными словами, хотим определить знак ее импульса $k$ [1, 21]. В качестве прибора для такого измерения используем макроскопическое устройство: «шарик» массы $M\gg\mu$, расположенный на вершине усеченного конуса и способный свободно, без трения двигаться по поверхности конуса. Потенциальная энергия этого шарика $U(Q)$ как функция координаты его центра тяжести $Q$ приведена на рис. 7. На

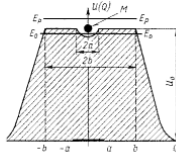

Рис. 7. Потенциальная энергия $U(Q)$ шарика $M$, расположенного на вершине усеченного конуса:
$Q$ — координата центра масс шарика. Указаны два уровня энергии $E_0$ и $E_P=P^2/2M$; $U_0$ — энергетическая высота конуса

вершине конуса имеется неглубокая ямка шириной $2a$. Энергия $\varepsilon$, необходимая для освобождения шарика из этой ямки, считается очень малой: $\varepsilon=E_0-E_0'\ll\ll\varepsilon_k$ — кинетической энергии частицы $\mu$ ($\varepsilon_k=\hbar^2k^2/2\mu$). В силу малости $\varepsilon$ шарик находится в неустойчивом равновесии, и после рассеяния на нем микрочастицы он движется направо или налево по вершине усеченного конуса. При достижении края вершины ($Q=\pm b$) шарик начнет падать вниз и может приобрести как угодно большую энергию $E_P=P^2/2M=U_0$ (здесь $P$ — импульс шарика, а $U_0$ — энергетическая высота конуса, которая может быть очень большой). Таким образом, микроскопическое явление — рассеяние частицы $\mu$ на шарике — порождает явление макроскопическое.

Мы будем считать температуру шарика равной 0 К и будем игнорировать его сложную структуру*. При этих предположениях исходное состояние шарика можно считать «чистым» и приближенно описывать его волновой функцией осциллятора:

$$\Phi_0(Q)=\exp(-Q^2/a^2)/\sqrt[4]{\pi a}. \quad (12.2)$$

_____
* Поскольку она явно несущественна в рассматриваемом процессе.

Ширину вершины конуса $b$ будем считать много большей ширины ямки $a$. Волновую функцию шарика после получения импульса $P$ в области $-b<Q<+b$ можно считать плоской волной:

$$\Phi_P(Q)=\exp(iPQ)/\sqrt{2\pi}. \quad (12.3)$$

Ниже будет показано, что после рассеяния частицы $\mu$ на шарике он оказывается в двух неинтерферирующих между собой состояниях. Одно из них отвечает движению шарика направо ($P>0$), а другое — движению налево ($P<0$) в зависимости от того, в каком состоянии рассеялась на шарике микрочастица $\mu$: имея импульс $k$ или $-k$. Поэтому движение шарика по вершине конуса выполняет функцию анализатора, а дальнейшее падение его вниз — функцию детектора. Таким образом, в нашем простом примере прибор, как это и должно быть, выполняет обе функции, характерные для измерительных приборов.

Обратимся теперь к математической теории этого прибора. Предположим энергию взаимодействия шарика и микрочастицы в простейшем виде:

$$\hat{W}_{AB}=g\delta(Q-x) \text{ для } t>0; \quad (12.4)$$
$$\hat{W}_{AB}=0 \quad \text{для } t<0,$$

где $g$ — константа взаимодействия, которую можно считать малой. Это позволит нам пользоваться теорией возмущений. Обозначим волновую систему «частица — прибор» в начальный момент времени $\Psi_0(x, Q)$:

$$\Psi_0(x, Q)=\Phi_0(Q)\varphi_k(x). \quad (12.5)$$

Волновую функцию системы $\Psi(x, Q, t)$ для $t>0$ будем искать, исходя из уравнения Шредингера

$$i\hbar\partial\Psi/\partial t-\{\hat{H}_A(x)+\hat{H}_B(Q)+\hat{W}_{AB}(x, Q)\}\Psi=0, \quad (12.6)$$

здесь $\hat{H}_A(x)$, $\hat{H}_B(Q)$ — гамильтонианы свободного движения частицы $\mu$ и шарика $M$. Положим

$$\Psi(x, Q, t)=\Psi_0(x, Q)\exp[-i(E_0+\varepsilon_k)t]+u(x, Q, t), \quad (12.7)$$

где $u$ — рассеянная волна. Подставляя (12.7) в (12.6) и пренебрегая произведением $\hat{W}u$ (величиной порядка $g^2$), получаем уравнение для рассеянной волны

$$i\hbar \partial u(x, Q, t)/\partial t - \{\hat{H}_A(x) + \hat{H}_B(Q)\}u(x, Q, t) =$$
$$= W_{AB}(x, Q)\Psi_0(x, Q, t)\exp[-i(E_0+\varepsilon_0)t]. \quad (12.8)$$

Эту волну будем искать в виде разложения

$$u(x, Q, t) = \int \Phi_P(Q)\exp[-iE_P t/\hbar]u_P(x, t)\,dP. \quad (12.9)$$

Подставляя теперь (12.9) в (12.8), умножая полученное уравнение на $\Phi_{P'}^*(Q)$, интегрируя по $Q$ и пользуясь условием ортогональности:

$$\int \Phi_{P'}^*(Q)\Phi_P(Q)\,dQ = \delta(P'-P), \quad (12.10)$$

получаем уравнение для

$$i\hbar \partial u_P(x, t)/\partial t - \hat{H}_A(x)u_P(x, t) =$$
$$= g\Phi_P^*(x)\Phi_0(x)\varphi_k(x)\exp[i(E_P-\varepsilon_b)/\hbar - i\omega_0 t/\hbar], \quad (12.11)$$

здесь

$$\hbar\omega_k = \hbar^2 k^2/2\mu; \quad \hat{H}_A(x) = -(\hbar^2/2\mu)\,d^2/dx^2.$$

Уравнение (12.11) разрешим с помощью функции Грина $\mathcal{G}(x-x', t-t')$ свободного уравнения, т. е. уравнения (12.11) при $g=0$:

$$u_P(x, t) = \int_0^t \mathcal{G}_\mu(x-x', t-t')\,dt'dx'\rho(x', t')\,dx'dt', \quad (12.12)$$

где согласно (12.11) источник $\rho(x', t')$ равен

$$\rho_P(x', t') =$$
$$= g\Phi_P^*(x)\Phi_0(x)\varphi_k(x)\exp[i(E_P-E_0)t/\hbar - i\omega_0 t]. \quad (12.13)$$

Амплитуды $A^+$, $A^-$ входят в (12.13) линейно, поэтому достаточно вместо $\varphi_k(x)$ рассматривать функцию $A^\pm\exp(\pm ikx)$ и соответствующие рассеянные волны $u_P^\pm(x, t)$.

Как известно, функция $\mathcal{G}(x, t)$ для свободного движения частицы с массой $\mu$ равна [21]

$$\mathcal{G}_\mu(x, t) = \sqrt{\frac{\mu}{2\pi i\hbar}}\exp(i\mu x^2/2\hbar t)\frac{1}{\sqrt{t}}. \quad (12.14)$$

64

Подставляя теперь $u_P(x, t)$ в (12.9), мы видим, что в (12.9) войдет интеграл

$$\mathcal{G}_M(Q-x', t-t') = \int dP\Phi_P(Q)\Phi_{P'}^*(x')\exp[iE_P(t-t')] \quad (12.15)$$

который есть не что иное, как функция Грина для свободного движения частицы $M$ (шарика). Поэтому

$$u^\pm(x, Q, t) = gA^\pm \int_0^t \mathcal{G}_\mu(x-x', t-t')\mathcal{G}_M(Q-x', t-t')\,dx'dt'\Phi_0(x')\exp(\pm ikx')\exp[-iE_0 t'/\hbar] \quad (12.16)$$

или в раскрытом виде, полагая $\tau = t-t'$:

$$u^\pm(x, Q, t) = gA^\pm\exp(i\omega_0 t)\sqrt{\frac{\mu}{2\pi i\hbar}}\sqrt{\frac{M}{2\pi i\hbar}}\int_{-\infty}^{\infty}dx'\times$$
$$\times\int_0^t \frac{d\tau}{\tau}\exp\left[i\mu\frac{(x-x')^2}{2i\hbar\tau} + iM\frac{(Q-x')^2}{2i\hbar\tau}\right]\exp(\pm ikx')\times$$
$$\times\Phi_0(x')\exp(-i\omega_0\tau). \quad (12.17)$$

В этом интеграле можно выполнить интегрирование по $dx'$ (см. дополнение 9). Тогда получим

$$u^\pm(x, Q, t) = gA^\pm\exp(i\omega_0 t)\frac{\sqrt{\tau M}}{2\pi i\hbar}\sqrt{\pi}\int_0^t\frac{d\tau}{\tau}\exp(-i\omega_0\tau)\times$$
$$\times\frac{1}{A(\tau)}\exp[-B_\pm^2(\tau) + C(\tau)], \quad (12.18)$$

причем

$$A = -\frac{i}{a^2} - \frac{i}{2\hbar\tau}(\mu+M),\quad B_\pm = \frac{1}{2A}\left\{\mp k + \frac{1}{\hbar\tau}(\mu x + MQ)\right\},$$
$$C = \frac{i}{2\hbar\tau}(\mu x^2 + MQ^2). \quad (12.19)$$

Если теперь использовать исходное предположение $M\gg\mu$, то $\mathrm{Re}\,B_\pm^2$ принимает вид

$$\mathrm{Re}\,B_\pm^2 = [Q\pm v\tau]^2/a^2, \quad (12.20)$$

65

здесь $v = \hbar k/M$ — скорость шарика (см. дополнение 9).

Отсюда следует, что функция $u^+(Q, t)$ сосредоточена в области положительных $Q$: $0<Q<vt$, а функция $u^-(Q, t)$ сосредоточена в области отрицательных $Q$: $-vt<Q<0$, так что $u^+(Q, t)u^-(Q, t)\approx 0$. Дальнейшее движение шарика $M$ в область $Q>|b|$ практически будет совпадать с классическим движением падающего с «горки» шарика.

Таким образом, доказывается, что при движении шарика в области $-b<Q<b$ уничтожается интерференция состояний, принадлежащих различным направлениям движения шарика. По достижении области $Q>|b|$ шарик будет падать направо или налево, набирая как угодно большую энергию $U_0$. Этим восполняется детекторная функция нашего измерительного устройства: по падению шарика справа или слева мы узнаем знак импульса частицы $\mu$.

### ЛЕКЦИЯ 13.
### ТЕРМОДИНАМИЧЕСКИ НЕУСТОЙЧИВЫЙ ДЕТЕКТОР

В этой лекции рассматривается измерительный прибор, предназначенный для определения направления спина микрочастицы $\mu$ [21]. Предполагается, что эта частица обладает магнитным моментом $\mu$:

$$\mu = \mu_0\sigma, \quad (13.1)$$

где $\sigma$ — матрица Паули. Пучок частиц $\mu$ будем считать неполяризованным. Поэтому исходный ансамбль некогерентен и описывается статистическим оператором $\rho$, имеющим матричные элементы:

$$\hat{\rho}(x, x') = P_1\psi_1(x)\psi_1^*(x') + P_2\psi_2(x)\psi_2^*(x'), \quad (13.2)$$

где волновая функция $\psi_1$ — состояние частицы $\mu$ с проекцией спина на ось $OZ$, равной $+1/2$, а волновая функция $\psi_2$ представляет состояние с проекцией спина на $OZ$, равной $-1/2$. Эти состояния равновероятны, так что $P_1 = P_2 = 1/2$.

С помощью неоднородного магнитного поля, параллельного оси $OZ$, пучки $\psi_1$ и $\psi_2$ можно разделить пространственно так, что каждый направляется в

66

свой детектор $D_1$ или $D_2$ (см. рис. 5). Этим выполняется первая функция прибора — функция анализатора. Эта функция в рассматриваемом случае тривиальна, и мы ее рассматривать даже не будем, а сосредоточимся исключительно на работе детекторов. Достаточно рассмотреть один из них.

В качестве детектора рассмотрим макроскопическое собрание осцилляторов, обладающих магнитным моментом, которое находится в термодинамически неустойчивом состоянии. Магнитный момент осцилляторов $\mathfrak{M}$ можно выразить через заряд $e$ и механический момент $\mathbf{M}$ согласно известной формуле $\mathfrak{M} = e\mathbf{M}/2mc$, где $m$ — масса осциллятора; $c$ — скорость света.

Энергия взаимодействия частицы $\mu$ с одним из осцилляторов детектора будет равна

$$\mathscr{W} = -\mathfrak{M}\mathbf{H}(x). \quad (13.3)$$

В этой формуле $\mathbf{H}(x)$ — магнитное поле, воздействующее на осциллятор со стороны частицы $\mu$. Это поле выражается формулой

$$\mathbf{H}(x) = \mu/R^3 - \mathbf{R}(\mu\mathbf{R})/R^5. \quad (13.4)$$

В этой формуле $R$ — расстояние от частицы $\mu$ до осциллятора. Для нашей цели достаточно ограничиться рассмотрением двумерной задачи. Будем считать, что осцилляторы образуют в плоскости $xy$ двумерный кристаллик размером $a$. Далее будем считать, что длина волны частицы $\mu\lambda\gg a$.

Предположим, что частица поляризована в направлении оси $OZ$. При этих условиях скалярное произведение $(\mu\mathbf{R})$ в (13.4) равно нулю. Условие $\lambda\gg a$ позволяет заменить величину $1/R^3$ в (13.4) на ее среднее значение, так что $1/R^3 = 1/a^3$. При этих предположениях энергия взаимодействия частицы и $s$-го осциллятора принимает простой вид:

$$\mathscr{W}_s = \pm i\hbar\omega\partial/\partial\varphi_s, \quad (13.5)$$

где частота $\omega = \frac{1}{a^3}\frac{e}{2mc}\mu_0$; $i\hbar\partial/\partial\varphi_s = \hat{M}_{zs}$ — оператор проекции механического момента $s$-го осциллятора на ось $OZ$; $\varphi_s$ — полярный угол. Знаки отвечают двум возможным ориентациям спина частицы $\mu$. Достаточно ограничиться одной из возможностей.

67

**Appendix B**. Excerpt from the 2007 "Under the Sign of Quantum" by L. I. Ponomarev

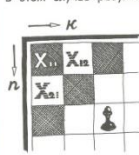